\documentclass[twocolumn,prb,showpacs,footinbib]{revtex4}
\usepackage{dcolumn}
\usepackage{bm}
\usepackage{enumerate}
\usepackage{graphicx,graphics,color,epsfig}
\usepackage{amsmath,amssymb}
\renewcommand{\vec}[1]{\boldsymbol{#1}}

\begin{document}

\title{The Haldane model under nonuniform strain}
\author{Yen-Hung Ho}
\affiliation{Department of Physics, National Tsing Hua University, Hsinchu City, Taiwan}
\affiliation{National Center for Theoretical Sciences (NCTS), Hsinchu City, Taiwan}
\author{ Eduardo V. Castro}
\affiliation{CeFEMA, Instituto Superior Tecnico, Universidade de Lisboa, Av. Rovisco Pais, 1049-001 Lisboa, Portugal}
\affiliation{Beijing Computational Science Research Center, Beijing 100084, China}
%\author{Maria A. H. Vozmediano}
%\affiliation{Instituto de Ciencia de Materiales de Madrid, and CSIC, Cantoblanco, 28049 Madrid, Spain}
\author{Miguel A. Cazalilla}
\thanks{Author to whom correspondence should be addressed: miguel.cazalilla@gmail.com}
\affiliation{Department of Physics, National Tsing Hua University, Hsinchu City, Taiwan}
\affiliation{National Center for Theoretical Sciences (NCTS), Hsinchu City, Taiwan}
\affiliation{Donostia International Physics Center (DIPC), 
Manuel de Lardizabal, 4 San Sebastian, Spain}

\begin{abstract}
 We study the Haldane model under strain using a tight-binding approach, and compare the obtained results with the continuum-limit approximation. As in graphene, nonuniform strain leads to a time-reversal preserving pseudo-magnetic field that induces (pseudo) Landau levels. Unlike a real magnetic field, strain lifts the degeneracy of the zeroth pseudo Landau levels at different valleys. Moreover, for the zigzag edge under uniaxial strain, strain removes the degeneracy within the pseudo-Landau levels by inducing a tilt in their energy dispersion. The latter arises from next-to-leading order corrections to the continuum-limit Hamiltonian, which are absent for a real magnetic field. We show that, for the lowest pseudo-Landau levels in the Haldane model, the dominant contribution to the tilt is different from graphene. In addition, although strain does not strongly modify the dispersion of the edge states, their interplay with the pseudo-Landau levels is different for the armchair and zigzag ribbons. Finally, we study the effect of strain in the band structure of the Haldane model at the critical point of the topological transition, thus shedding light on the interplay between non-trivial topology and strain in quantum anomalous Hall systems.  
\end{abstract}


\maketitle

\section{Introduction}

 The Haldane model~\cite{Haldane88} (HM) is a prominent representative of a wider class of lattice models of 2D quantum anomalous Hall (QAH) insulators that capture the essence of the quantum Hall effect (QHE). The QHE is characterized by a nontrivial topological invariant, the Chern number, which also characterizes the band structure of the HM. A time-reversal invariant generalization of the HM was introduced by Kane and Mele,~\cite{KaneMele} who noticed that the complex next-nearest neighbor hopping in the HM (referred to below as ``Haldane mass'') can be related to the intrinsic spin-orbit coupling (SOC) in graphene. The resulting Kane-Mele model provides a minimal description for the quantum spin Hall (QSH) effect, which has recently attracted much experimental and theoretical attention.~\cite{Bernevig,Hasan2010rmp,Konig} Currently, the search of experimental realizations of the Kane-Mele model based on enhancing the SOC in graphene is also a very active research field.~\cite{Weeks,Cresti,Brey,Garcia} As for the Haldane model, no solid-state realizations are available to date, although the model has been recently realized using ultracold atoms in optical lattices.~\cite{Jotzu14}

\begin{figure}
\centering
\includegraphics[scale=0.48]{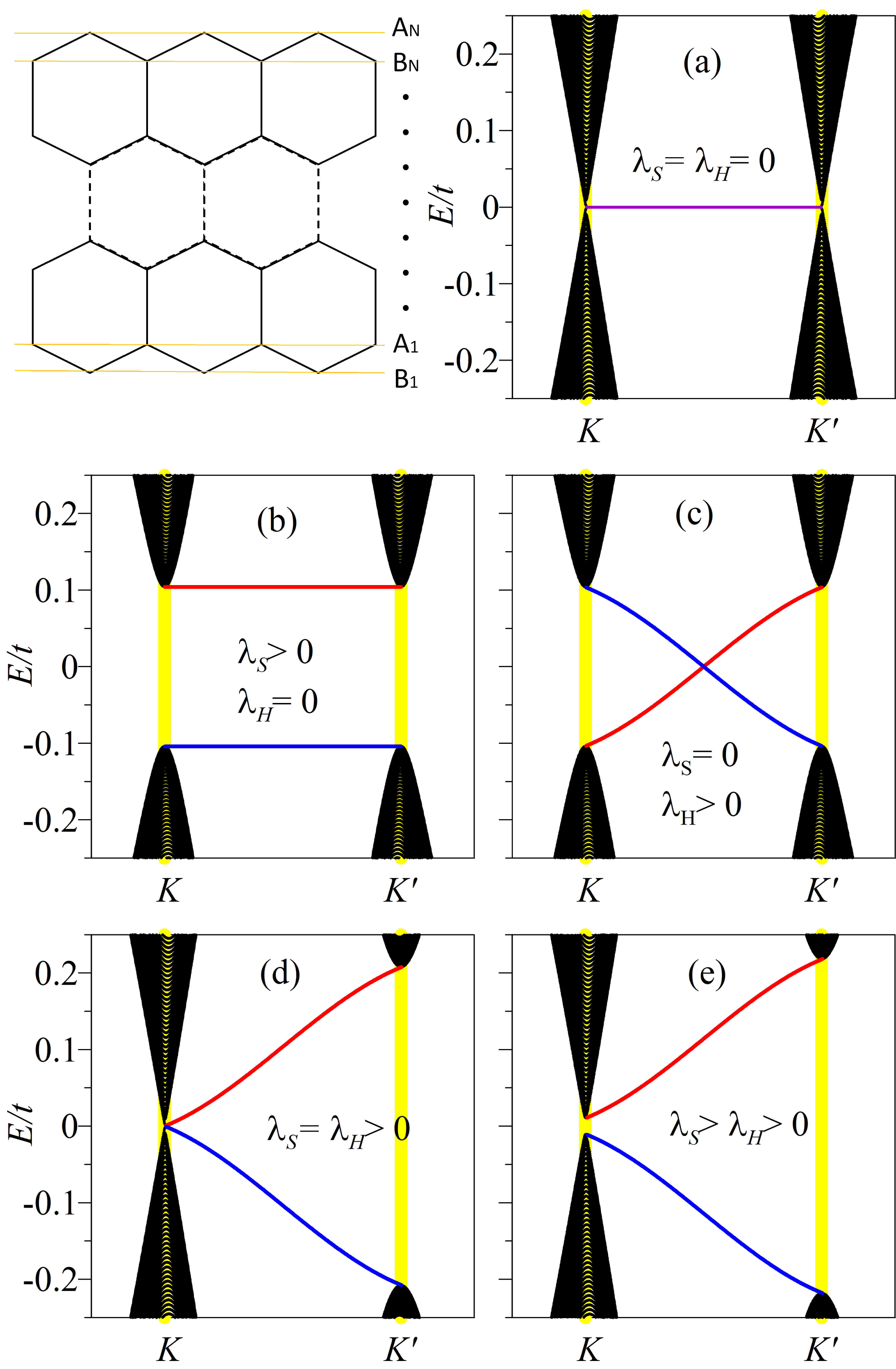}
\caption{(Color online) Ribbon geometry (we take the number of atomic rows $N=500$ for both the zigzag and armchar ribbons). Ribbon band structure for (a) pristine graphene (b) graphene with sublattice-inversion symmetry breaking Semenoff mass, $\lambda_S=3\sqrt{3}t^{\prime}$ (c) with Haldane hopping (Haldane mass  $\lambda_H=3\sqrt{3}t'$) (d) with both Semenoff and Haldane masses, $\lambda_S=\lambda_H=3\sqrt{3}t^{\prime}$ and (e) $\lambda_S>\lambda_H=3\sqrt{3}t^{\prime}$, where $t^{\prime}=0.02t$. The red and blue curves show the topological edge states localized at the top and bottom edges, respectively. The vertical yellow stripes mark the position of $K$ and $K'$ points.}
\label{fig:fig1}
\end{figure}

\begin{figure}
\centering
\includegraphics[scale=0.48]{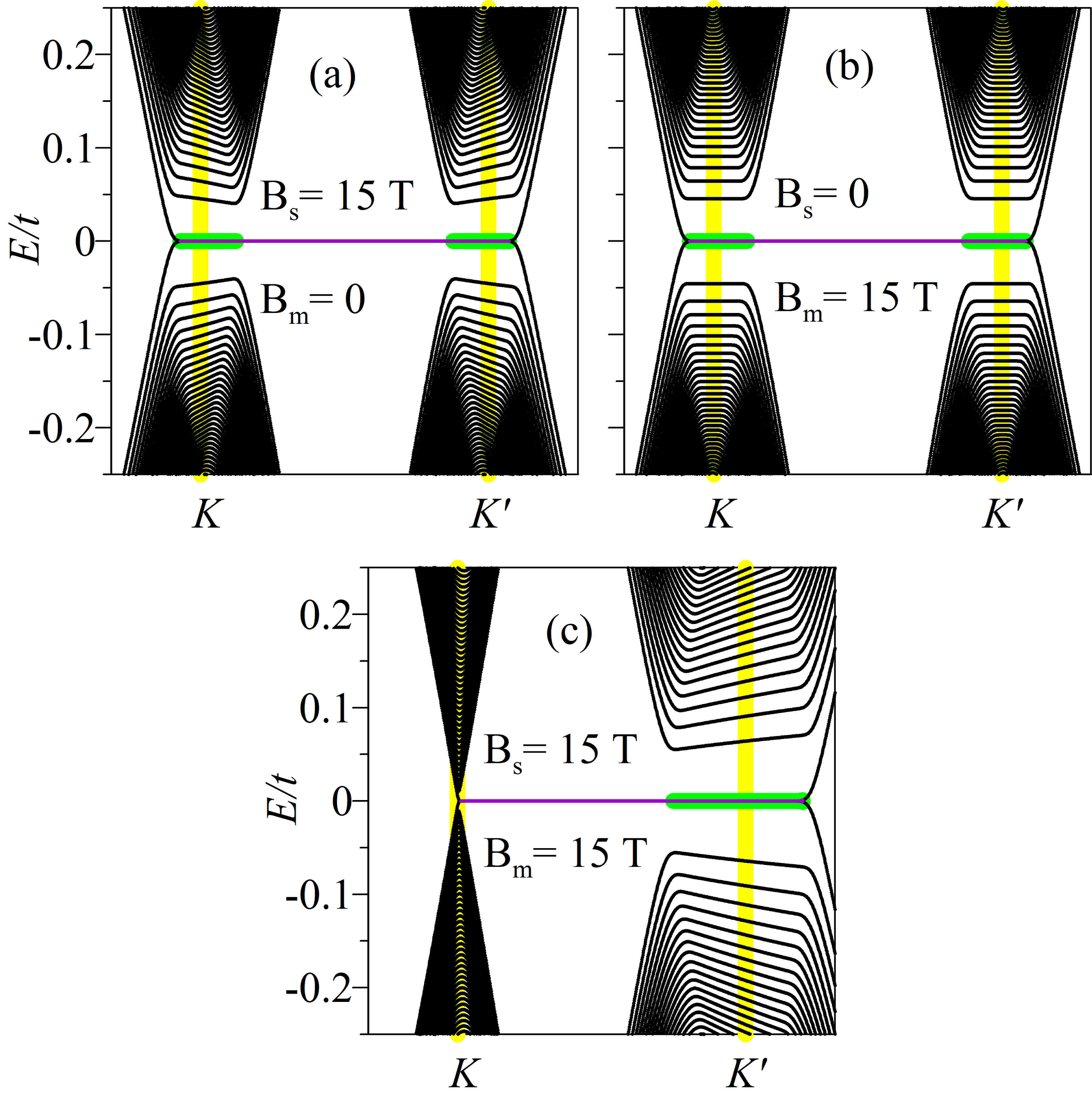}
\caption{(Color online) Band structure of graphene zigzag ribbon (a) under nonuniform strain corresponding to a pseudo-magnetic field $\mathcal{B}_s$=15 T (b) under a \emph{real} magnetic field $\mathcal{B}_m$=15 T (c) under both nonuniform strain and magnetic field: $\mathcal{B}_s=\mathcal{B}_m$= 15 T. The green colored subbands indicate the zeroth LLs.}
\label{fig:fig2}
\end{figure}

 In the presence of a potential that breaks sublattice inversion symmetry (henceforth referred to as ``Semenoff mass''~\cite{Semenoff84}), the HM is also a paradigm for topological phase transitions.~\cite{Haldane88} The Semenoff mass  lifts the degeneracy between the two valleys and eventually it drives the closing of the gap of the topologically non-trivial phase, as shown in Fig.~\ref{fig:fig1}(d). As the gap reopens upon further increase of the sublattice inversion symmetry breaking potential, the Chern number characterizing the bands of the topological phase vanishes.~\cite{Haldane88} In fact, it is interesting to notice that a similar phenomenon takes place for nonuniformly strained graphene in a magnetic field: Strain induces a time-reversal preserving pseudo-magnetic field, which may annhilate the real magnetic field at one valley, thus driving the collapse of the gap between Landau levels (LLs)~\cite{Zhai10,Chaves10} as shown in Fig.~\ref{fig:fig2}, and therefore the destruction of the Chern invariant.

Rather than studying the possible modifications of the Chern invariant due to nonuniform strain, here we adopt a more  practical approach. By relying on the bulk-edge correspondence, we study the effects of strain on the dispersion of the edge states for non-uniformly strained ribbons. We analyze the effects of strain for the two canonical honeycomb lattice terminations, namely the zigzag and armchair edges. As shown below, nonuniform strain has different effects on these two kinds of edges.

 The possibility of engineering the electronic properties of two-dimensional materials like graphene~\cite{Suzuura02,Guinea10,Low10,Ghaemi13,xianpeng17,castro17} and transition metal dichacogendides~\cite{cog_tmdc,Castro16} using strain (``straintronics'') is currently a very active research field.~\cite{Amorim16} Interestingly, to the best of our knowledge, the question of how nonuniform strain affects the topological features of the HM and the topological phase transition has received little attention so far. Ghaemi \emph{et al.}~\cite{Ghaemi13} briefly considered the effect of strain in the HM in their work. However, they relied on a continuum-limit (i.e. Dirac equation) description, which is not accurate away from the Dirac points. In this work, we use a tight-binding approach~\cite{Ho11} to show that the effects of strain on the topological gap are different from those of a Semenoff \emph{mass}. Tight-binding calculations are necessary in order to assess the accuracy of the continuum approximation and to ascertain other effects of nonuniform strain well away from the Dirac points. Within the tight-binding approach we have found that strain induces a tilt of the pseudo Landau levels (pLLs). For the dominant contribution to this tilt, the lowest pLLs is different from those in strained graphene. Strain can also have a weak effect on the magnitude of the band gap. However, it does not strongly  modify the dispersion of the edge states neither on the zigzag nor armchair edge. This shows that there is no competition between strain and the Haldane mass, unlike the case of the Semenoff mass. Indeed, the topological phase transition can be regarded as a competition between the QAH effect, induced by the Haldane \emph{mass}, and the quantum valley Hall effect (QVHE), induced by the Semenoff mass. In this regard, it is worth noting that (nonuniform) strain also drives the QVHE in graphene.~\cite{Guinea10} However, as shown below,  the strain-induced QVHE does not compete with the QAH induced by the Haldane mass. 

The rest of this article is organized as follows. In Sec.~\ref{sec:model}, we introduce the Haldane model and briefly review the most relevant results about it. We first consider the unstrained model (cf. Sec.~\ref{sec:hm}), before addressing the effects of nonuniform strain using the continuum-limit approach. In the same section, we also discuss the tilt of the pLLs arising from next-to-leading order corrections. In Sec.~\ref{sec:tight}, we take up the study of the lattice HM as realized in zigzag and armchair ribbons under nonuniform strain. The effects of strain on the topological transition are described in Sec.~\ref{sec:tt}. Finally, in Sec.~\ref{sec:concl} we offer the conclusions for this work. For the sake of completeness, we have included some details of the calculations in the Appendices.

\section{Model and its continuum limit}\label{sec:model}

 In this section, we review some of the most important results for the HM. We also review its continuum-limit description both in the presence and absence of strain.
 
\subsection{Model}\label{sec:hm}

The Haldane model (HM) describes the hopping of \emph{spinless} electrons (interactions are neglected) on a honeycomb lattice. The Hamiltonian for HM reads
\begin{align}
\mathcal{H}&=\mathcal{H}_0 +\mathcal{H}_T +\mathcal{H}_S,\label{eq:hm1} \\
\mathcal{H}_0 &= - \sum_{i,j} t_{ij}\left[ a^{\dag}_i b_{j} + \mathrm{h.c.}\right] ,\label{eq:hm2} \\
\mathcal{H}_T &= -  \sum_{i,j} t^{\prime}_{ij} \left[  e^{i\nu_{ij}\psi}
a^{\dag}_i  a_{j} +  e^{i\nu_{ij}\psi}  b^{\dag}_i b_{j}  \right],
\label{eq:hm3} 
\end{align}
where $t_{ij} = t$ for $i$ and $j$ nearest neighbors (and zero otherwise), and $t^{\prime}_{ij} = t^{\prime} (= t/50 $ in this work) for $i$ and $j$ next nearest neighbors (and zero otherwise). In the above expression, $a_i^{\dag}$ corresponds to a creation operator for an electron at site $i$ on the $A$ sublattice and $b_i^{\dag}$ is a creation operator for an electron on the $B$ sublattice. Note that the second nearest neighbor hoping is complex for $\psi\neq 0$, where $\nu_{ij}=\mathrm{sgn}({\bf d}_1\times{\bf d}_2)_z=\pm 1$.~\cite{Haldane88,Liu15} Here ${\bf d}_{1,2}$ are the vectors along the two bonds linking the second nearest neighbors. In Eq.~\eqref{eq:hm1}, $\mathcal{H}_S$ describes the Semenoff mass, which breaks the sublattice inversion symmetry~\cite{Semenoff84,Liu11,Tabert13,Ezawa12,Barkhofen13} and takes the form:
\begin{equation}
\mathcal{H}_S = \epsilon_0 \sum_{i} \left[ a^{\dag}_i a_i  - b^{\dag}_{i}b_i\right]. 
\end{equation}
In the absence of this term (i.e. for $\epsilon_0 = 0$) and for $\psi = \tfrac{n\pi}{2}$, with $n$ an odd integer, the HM exhibits particle-hole symmetry. 

 In order to describe strain in the tight-binding approach, the hoping amplitudes $t$ and $t^{\prime}$ in the HM (Eqs.~\eqref{eq:hm1} and \eqref{eq:hm2}) are shifted according to:
\begin{align}
t_{ij} &\to t_{ij} + \frac{\beta t}{a^2} \left( \vec{\mathcal{R}}_i
- \vec{\mathcal{R}}_j\right) \cdot \left[\vec{u}(\vec{\mathcal{R}}_i) - \vec{u}(\vec{\mathcal{R}}_j) \right],\label{eq:st} \\ 
t^{\prime}_{ij} &\to t^{\prime}_{ij} + \frac{\beta^{\prime} t^{\prime}}{3 a^2} \left( \vec{\mathcal{R}}_i
- \vec{\mathcal{R}}_j\right) \cdot \left[\vec{u}(\vec{\mathcal{R}}_i) - \vec{u}(\vec{\mathcal{R}}_j) \right],\label{eq:stp} 
\end{align}
where $i,j$ correspond to the nearest neighbors in the first line and to the next nearest neighbors in the second line; $\vec{\mathcal{R}}_j$ stands for the spatial position of site $j$, $\vec{u}(\vec{r})$ is the atomic displacement field, and $a$ is the C-C distance. The Gr\"uneisen parameters $\beta = \tfrac{d\log t}{d \log a}\simeq -3$, as  for graphene~\cite{Amorim16,Goerbig11,Guinea10, Neto09,deJuan13,Suzuura02,Low10} and we set $\beta^{\prime} = \tfrac{d\log t^{\prime}}{d \log \sqrt{3}a}\simeq -1$.

In addition, with the application of a real magnetic field, $\vec{\mathcal{B}}_m$, the vector gauge field $\vec{\mathcal{A}}_m$ yields a Peierls phase which modifies the hopping amplitudes by a factor~\cite{Ho11} $e^{{i2\pi\over \phi_0}\Delta G(\vec{\mathcal{R}}_j,\vec{\mathcal{R}}_k)}$, where 
\begin{eqnarray}
\cr&&\Delta G(\vec{\mathcal{R}}_j,\vec{\mathcal{R}}_k)=
\cr&&\int^1_0(\vec{\mathcal{R}}_k-\vec{\mathcal{R}}_j)\cdot\vec{\mathcal{A}}_m\left(\vec{\mathcal{R}}_j+\lambda(\vec{\mathcal{R}}_k-\vec{\mathcal{R}}_j)\right)d\lambda.
\end{eqnarray}
Here we choose the gauge $\vec{\mathcal{A}}_m=(-\mathcal{B}_my,0)$ for zigzag and $\vec{\mathcal{A}}_m =(0,\mathcal{B}_mx)$ for armchair. Both result in a perpendicular magnetic field $\vec{\mathcal{B}}_m=\vec{\nabla}\times\vec{\mathcal{A}}_m =\mathcal{B}_m \boldsymbol{\hat{z}}$. 

\subsection{Continuum description (unstrained system)}

 In the continuum limit, the electronic states in the neighborhood of the $K$ ($K^{\prime}$) points of Brillouin zone can be described by the following Hamiltonian (henceforth we work in $\hbar = 1$ units):
\begin{equation}
\mathcal{H} = v_F \left(\tau\sigma_x  p_x + \sigma_y p_y \right) - \lambda_H \sigma_z\tau +  \lambda_S \sigma_z,\label{eq:kpham}
\end{equation}
where $v_F = 3 at/2$ is the Fermi velocity, $\lambda_H = 3\sqrt{3}t'\sin\psi$ and $\lambda_S =\epsilon_0 $. In the above expression, the Pauli matrices $\sigma_x, \sigma_y, \sigma_z$ describe the sublattice pseudo-spin, and $\tau = +1$ ($\tau = -1$) for the states around the $K$ ($K^{\prime}$) point.

 In the neighborhood of $K$ and $K^{\prime}$, the spectrum of bulk states exhibits a gap which has the form:
\begin{equation}
\epsilon_{\lambda\tau}(\vec{p})= s \sqrt{v^2_F |\vec{p}|^2 + \Delta^2_{\tau} }, \label{eq:dirac}
\end{equation}
where $s = +1$ ($s = -1$) for the conduction (valence) band, and $\Delta_{\tau}= -\tau\lambda_H+\lambda_S$. The Chern number is determined by 
\begin{equation}
\nu=\tfrac{1}{2}[\mathrm{sgn}(\Delta_-)-\mathrm{sgn}(\Delta_+)].
\end{equation}
Thus, for $\lambda_{H,S} > 0$ and $\tau = +1$ (i.e. the $K$ point), the bulk gap closes for $\lambda_H = \lambda_S$. At such a point, the system becomes a metal, as shown in Fig.~\ref{fig:fig1}(d). For $\lambda_S < \lambda_H$, the system is adiabatically connected to the HM with $\lambda_S = 0$, which is a Chern insulator with topologically-protected gapless edge states. On the other hand, for  $\lambda_S > \lambda_H$, the system is a \emph{trivial} insulator for which the Chern number vanishes (i.e. $\nu = 0$) and no topologically protected edge states exist.

\subsection{Continuum description (strained system)}\label{sec:strainhm}

 To leading order, the effect of nonuniform strain on the honeycomb lattice can be described by a pseudo-gauge field $\vec{\mathcal{A}}$, which accounts for the shift in the foci of the Dirac cones at the $K$ and $K^{\prime}$ valleys. Thus, Eq.~\eqref{eq:kpham} becomes:
\begin{align}
\mathcal{H} &=v_F\left[\tau\sigma_x(p_x+ g \mathcal{A}_x)+\sigma_y(p_y+ g\mathcal{A}_y)\right] \notag \\
&-\lambda_H\sigma_z\tau+\lambda_S\sigma_z, \label{eq:kp}
\end{align} 
where $\vec{\mathcal{A}} =  \tfrac{{3\beta}}{4a}(u_{xx}-u_{yy},-2u_{xy})\tau$, where $u_{\alpha\beta} = \frac{1}{2}\left(\partial_{\alpha} u_{\beta} + \partial_{\beta} u_{\alpha} \right)$ is the strain tensor and $g = \tfrac{2}{3}$. Corrections to this Hamiltonian, beyond the leading order, which have previously been classified using symmetry arguments,~\cite{deJuan13} are discussed below and in Appendixes~\ref{app:modgap} and ~\ref{app:tilt}.
 
 Nonuniform strain leads to a pseudo-gauge and pseudo-magnetic field. The bulk spectrum of the continuum Hamiltonian is introduced at the outset of this section, and the position of the zeroth pLLs (henceforth referred to as 0th pLLs) is computed in the Appendix~\ref{app:0ll}. It is shown there that, compared to the case of a real magnetic field, the position of the 0th pLLs is reversed. This is confirmed by the tight-binding calculations, as we discuss as follows.

 The continuum description is only accurate in the neighborhood of the $K$ and $K^{\prime}$ points. In order to obtain a more complete picture of the effects of nonuniform strain in the HM, the tight-binding approach is necessary (see Sec.~\ref{sec:tight}). To study the edge states, we consider the ribbon geometry, whose band structure can be obtained using numerical diagonalization. Preserving translational invariance in the direction along the ribbon edge allows us to work with smaller matrix sizes. Requiring translational invariance along the ribbon edge constrains the kind of nonuniform strain that we can study. Thus, for a displacement field $\vec{u}(\vec{r})$, taking the ribbon edge along  $\hat{\vec{n}}$ ( $\hat{\vec{n}}^2 = 1$), requires the atomic displacement along the edge, i.e. $u_{\parallel} = \hat{\vec{n}}\cdot \vec{u}$, to be a constant (which we  take to be zero in what follows). In addition, we must require that the displacement perpendicular to the edge, i.e. $u_{\perp} =\left|\hat{\vec{z}}\cdot\left(\vec{\hat{n}}\times \vec{u}\right)\right|$ ($\vec{\hat{z}}$ is the unit vector perpendicular to the ribbon) is independent of the coordinate along the edge, i.e. $\vec{\hat{n}}\cdot \vec{r}$. 

 In what follows, we specialize the above considerations to the two most general types of edge terminations of the honeycomb lattice, namely the zigzag and the armchair. First, let us consider a zigzag edge (cf. Fig.~\ref{fig:fig1}) along the $x$ direction (i.e. $\vec{\hat{n}}=\vec{\hat{x}}$). The armchair direction corresponds to the $y$-axis (cf. Fig.~\ref{fig:fig1}). For the zigzag edge, we require that $u_{\parallel} = u_x = 0$ and $u_{\perp} = u_{y} = u_y(y)$, which implies that the strain tensor components are  $u_{xx}= 0$ and $u_{xy} = 0$, and $u_{yy} = \partial_{y} u_{y}(y)$. In order to obtain a constant pseudo-magnetic field, we use a nonuniform strain of the form $u_{yy}=\partial_yu_y=2C(y-{L\over 2})$, where the ribbon width is $L={3\over 2}aN$. For this choice, the zigzag ribbon remains unstrained (clamped) at the center, but it is stretched at the top and compressed at the bottom edge. The strain gauge field is $\vec{\mathcal{A}} = - \tfrac{3\beta}{4a}u_{yy}(1,0)\tau$, which results in a pseudo magnetic field $\vec{\mathcal{B}}_s={{3\beta}\over{2a}}C\tau \vec{\hat{z}}$. Note that, unlike a real magnetic field, the pseudo-magnetic field has opposite sign at oposite valleys (i.e. $\vec{\mathcal{B}}_s \propto \tau$)  because it does not break time-reversal symmetry. 
 
 Notice that for the zigzag ribbon the strain tensor is uniaxial, that is, not traceless: $u_{xx}+u_{yy} = u_{yy} \neq 0$. In such cases,  together with the pseudo-gauge field, $\vec{\mathcal{A}}$, the following terms are also present in the continuum Hamiltonian in Eq.~\eqref{eq:kp}:
\begin{equation}
\mathcal{H}^{\prime} = \left(u_{xx}+u_{yy}\right) \left( \gamma\sigma_0 + \gamma^{\prime} \sigma_z + \gamma^{\prime\prime}\tau \sigma_z\right),
\label{eq:sgap}
\end{equation}
where $\sigma_0$ is the unit matrix in the sublattice pseudo-spin. For example, for the HM with the nearest and next to the nearest-neighbor hopping, $\gamma = \tfrac{3}{2}\beta^{\prime} t^{\prime} \cos\psi$ (see Appendix~\ref{app:modgap}), which vanishes for $\psi = \frac{n\pi}{2}$. The expressions for $\gamma^{\prime}$ and $\gamma^{\prime\prime}$ for the HM are given in the Appendix~\ref{app:modgap}. However, note that specific physical realizations of the HM will contain longer range hopping terms, which lead to modified values for $\gamma^{\prime}$, and $\gamma^{\prime\prime}$. In particular, the deformation potential $\propto \gamma \sigma_0$ has been shown to lead to the collapse of the Landau levels for large values of $\gamma$.~\cite{castro17} A similar collapse is also expected to take place for the HM. Here we show that the non-vanishing trace of the strain tensor introduces further corrections to the continuum Hamiltonian,~\cite{deJuan13} which account for the tilt of the pseudo-Landau levels (see Appendix~\ref{app:tilt} for detailed discussion of this point and Sec.~\ref{sec:tt} for a discussion of their effect on the tight-binding band structure).

 Beyond the leading order, the following corrections are obtained:~\cite{deJuan13}
\begin{align}
\mathcal{H}^{\prime\prime} =&\quad \alpha (u_{xx}+u_{yy})(\tau \sigma_x p_x+\sigma_y p_y)\notag\\
   &+ \alpha^{\prime}\left[\left( u_{xx}\tau \sigma_x p_x +  u_{yy} p_y \sigma_y \right) +  u_{xy}\left(\sigma_y p_x +  \tau \sigma_x p_y \right)\right],
  \label{eq:strain_veloc}
\end{align}
where $\alpha = \tfrac{3}{8}a\beta t$ and $\alpha^{\prime} = 2\alpha$ for the HM. The first term ($\propto \alpha$) is an isotropic correction to the magnitude of the Fermi velocity $v_F$,~\cite{deJuan13} whereas the second term ($\propto \alpha^{\prime}$) is an anisotropic correction to $v_F$.~\cite{deJuan13} 

 As we show in detail in Appendix~\ref{app:tilt}, Eqs.~\eqref{eq:sgap} and \eqref{eq:strain_veloc} lead to momentum dependent corrections to the pLL energy:
\begin{align}
\Delta \epsilon^{\prime}_{n\pm}(\tau) &= \langle \Psi_{n\pm}(\tau) | \mathcal{H}^{\prime} | \Psi_{n\pm}\rangle \propto \left(\beta^{\prime} t^{\prime}\right)  \left( p_x \ell\right) ,  \label{eq:tiltp}\\
\Delta \epsilon^{\prime\prime}_{n\pm}(\tau) &= \langle \Psi_{n\pm}(\tau) | \mathcal{H}^{\prime\prime} | \Psi_{n\pm}\rangle \propto \left( \beta t\right) (p_x a), \label{eq:tiltpp}
\end{align}
where $|\Psi_{n\pm}(\tau)\rangle$ are the pLLs eigenfuctions (see Appendix~\ref{app:0ll} for details) and $\ell = (2 g B_s)^{-1/2}$ is the pseudo-magnetic length. Explicit expressions for these energy corrections are given in Eqs.~\eqref{eq:corra} and \eqref{eq:corrb} in Appendix~\ref{app:tilt}. Notice that even though $\beta^{\prime} t^{\prime}/(\beta t) \simeq \tfrac{1}{150}\ll 1$, the continuum limit approximation requires that $\ell/a \gg 1$ and also 
$\Delta\epsilon^{\prime\prime}_{n\pm}$ becomes small for $n\to 0$ whereas $\Delta \epsilon^{\prime}_{n\pm}$  approaches a constant. Thus, we find that $\Delta \epsilon^{\prime}_{n\pm}(\tau)/\Delta \epsilon^{\prime\prime}_{n\pm}(\tau) \sim 10$, for the lowest pLLs. Thus, we find that the tilt arising from the deformation potential of Eq.~\eqref{eq:sgap} dominates over the correction arising from the Fermi velocity correction (cf. Eq.~\eqref{eq:strain_veloc}). Furthermore, when considering the 0th pLLs, we find $\Delta \epsilon^{\prime\prime}_{0}(\tau) = 0$ and $\Delta \epsilon^{\prime}_0(\tau) \neq 0$. Nevertheless, for graphene $\Delta \epsilon^{\prime}_{n\pm} = 0$, so that the tilt observed in the pLLs with $n > 0$, Fig.~\ref{fig:fig2}(a), arises entirely from the corrections to the Fermi velocity, Eq.~\eqref{eq:strain_veloc} (see also Eq.~\eqref{eq:corra} in the Appendix). Also, the tilt does not appear for a real magnetic field because no terms similar to $\mathcal{H}^{\prime}$ and $\mathcal{H}^{\prime\prime}$ are present in this case.  

 Using the expressions provided in the Appendix~\ref{app:tilt} for $\Delta \epsilon^{\prime}_{n\pm}(\tau)$ and $\Delta \epsilon^{\prime\prime}_{n\pm}(\tau)$, the following relations can be established: For the HM in the presence of Semenoff mass, the total energy shift, $\Delta \epsilon_{n\pm}(\tau) = \Delta \epsilon^{\prime}_{n\pm}(\tau) + \Delta \epsilon^{\prime\prime}_{n\pm}(\tau)$, obeys
\begin{equation}
\Delta \epsilon_{n\pm}(\tau)  = -\Delta \epsilon_{n\mp}(\tau).
\label{eq:rel1}
\end{equation}
That is, the tilts of the pLLs splitting off the conduction and valence band have  opposite signs. In addition, in the absence of Semenoff mass and for $\psi = \tfrac{n\pi}{2}$ which is relevant to the tight-binding calculations that we report in the following section,  we have
\begin{equation}
\Delta \epsilon_{n\pm}(\tau) = \Delta \epsilon_{n\mp}(-\tau),
\label{eq:rel2}
\end{equation}
which relates the tilts of the pLLs in opposite valleys. Combining Eqs.~\eqref{eq:rel1} and \eqref{eq:rel2},
we obtain:
\begin{equation}
\Delta \epsilon_{n\pm}(\tau) = -\Delta \epsilon_{n\pm}(-\tau).
\label{eq:rel3}
\end{equation}
This result implies that  pLLs with the same indices $(n,\pm)$  at opposite valleys are tilted in opposite directions. In section~\ref{sec:zigzag},  these results are confirmed numerically by diagonalizing  the tight-binding Hamiltonian of a  zigzag ribbon under uniaxial strain. It is also worth noting that, in a general case (e.g. $\psi \neq \tfrac{n\pi}{2}$ and non-zero $\gamma^{\prime}$), there will be extra contributions to energy shift, $\Delta \epsilon^{\prime}_{n\pm}$. The latter do not respect the last two of above relations, namely Eqs.~\eqref{eq:rel1} and \eqref{eq:rel2}, which are violated by the correction to the Semenoff mass, i.e. by the term proportional to $\gamma^{\prime}$ in Eq.~\eqref{eq:sgap}.

 For the armchair edge, translational invariance requires that $\partial_y u_{\parallel}=\partial_y u_{y} = 0$ and $\partial_y u_{\perp}=\partial_y u_{x}=0$. Thus, we choose $u_y=C(x-{L\over 2})^2$ where the ribbon width $L={\sqrt{3}\over 2}aN$. The strain tensor components are $u_{xx}=u_{yy}=0$ and $u_{xy}=C(x-{L\over 2})$, and therefore the strain in this case is pure shear strain as the strain tensor is traceless. Since the lattice displacement is along $y$, the ribbon is thus bent like a Corbino, with the maximum strain taking place at the right edge, vanishing at the ribbon center, and becoming negative (i.e. compression) at the left edge. This strain gives rise to a gauge field $\vec{\mathcal{A}}=-{{3\beta}\over{2a}}u_{xy}(0,1)\tau$ and thus $\vec{\mathcal{B}}_s=-{{3\beta}\over{2a}}C\tau\:\vec{\hat{z}}$.

\section{Tight-binding approach}\label{sec:tight}

In this section, we discuss the results obtained by studying the HM in strained ribbons using the tight-binding approach. We show that this approach, which yields the full band structure of the ribbon, confirms the results obtained from the continuum model, but it also provides detailed information about the dispersion of the edge states away from the $K$ and $K^{\prime}$ points. We first focus on the case for which the sublattice inversion-breaking potential (i.e. the Semenoff mass) is absent. In the following section, we discuss the effect of the combination of strain and Semenoff mass at the critical point of the topological phase transition.
 
\begin{figure}
\centering
\includegraphics[scale=0.48]{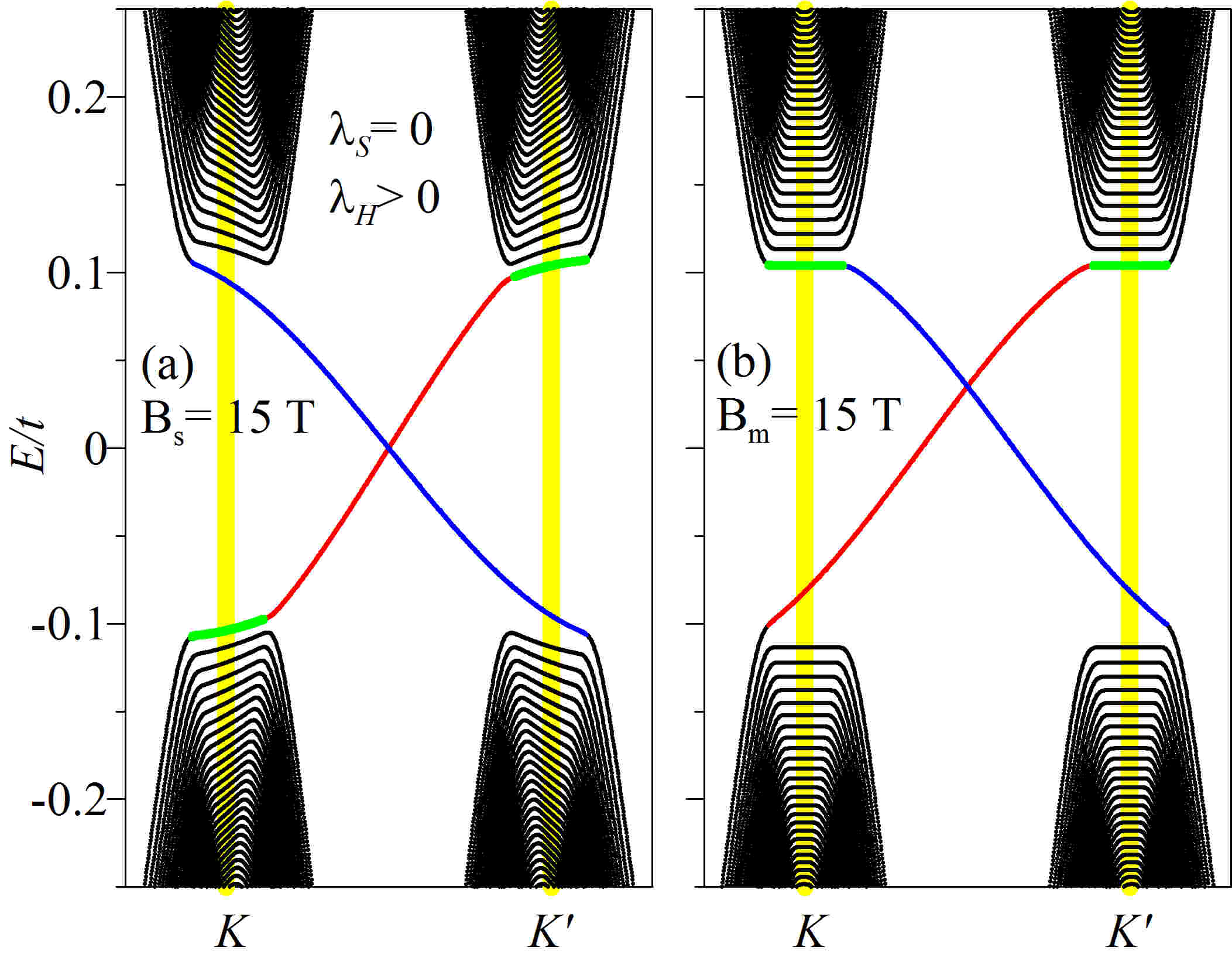}
\caption{(Color online) Band structure of a zigzag ribbon described by the Haldane model with a Haldane mass $\lambda_H=3\sqrt{3}t'$ subject to (a) a nonuniform  uniaxial strain leading to a constant pseudo-magnetic field $\mathcal{B}_s =15$ T and (b) a \emph{real} magnetic field $\mathcal{B}_m = 15$ T. In both cases, the position of the zeroth Landau levels is highlighted in green.}
\label{fig:fig3}
\end{figure}
\subsection{Zigzag ribbon}\label{sec:zigzag}
 In Fig.~\ref{fig:fig3}(a), we show the spectrum of a zigzag ribbon described by the HM under nonuniform (uniaxial) strain. For the zigzag direction, the (crystal) momentum along the edge ($k_x$) is a good quantum number and the $K$ and $K^{\prime}$ valleys are projected at opposite $k_x$ points relative to the center of ribbon Brillouin zone (BZ). As shown in Fig.~\ref{fig:fig3}(a), the position of the $0$th pLLs at opposite valleys $K$ and $K^{\prime}$ is symmetrically located in energy relative to the center of the band gap. This should be contrasted with the case of a real magnetic field shown in Fig.~\ref{fig:fig3}(b),~\cite{araujo14} for which the 0th LLs at the two valleys are degenerate in energy (the levels split off the bottom of the conduction band for our choice of model parameters). Notice that a perpendicular magnetic field preserves the 2D inversion symmetry (i.e. $x \to -x$ and $y \to -y$ but $z \to z$). In the absence of sublattice symmetry breaking, 2D inversion symmetry is responsible for the degeneracy of the bulk spectra in the HM at $K$ and $K^{\prime}$, of which the degeneracy of the 0th LL is just one manifestation. However, strain does not respect this degeneracy. This can be interpreted as an effective breaking of the inversion symmetry. Note as well that, unlike the inversion-symmetry breaking induced by the Semenoff mass which arises from a potential that varies rapidly on the scale of the lattice parameter,  nonuniform strain varies slowly on the scale of the lattice. 

Besides the different position of the 0th pLLs at opposite valleys, strain also introduces a tilt of the pLLs, as it can be seen Fig.~\ref{fig:fig3}(a). This tilt is absent for the ribbon subject to a perpendicular magnetic field. The tilt has also the opposite sign for the pLLs splitting off the conduction and valence bands. In addition, for the same $n$ and band, pLLs at opposite valleys are tilted in opposite directions. These observations are in agreement with the relations~\eqref{eq:rel1} through \eqref{eq:rel3}, derived in Sec.~\ref{sec:strainhm} from the continuum-limit approach. 
    
 Nonuniform strain slightly modifies the dispersion of  edge states, as shown in Fig.~\ref{fig:fig3}(a). However, despite the modification, the integrity of the topologically protected states on the zigzag edge is not affected by the application of uniaxial strain. Nevertheless, notice that, compared to the case of a real magnetic field, the bands of edge states corresponding to different edges disperse differently and are no longer symmetric about the center of the BZ. The top edge band connects the 0th pLLs and exhibits a different dispersion from the bottom edge band. Indeed, from an analysis of the wave functions, we observe that the latter band is localized at the bottom edge for all values of $k_x$ and does not mix with the pLLs. However, for the top edge, the edge states evolve into bulk states for $k_{x}$ in the neighborhood of $K$ and $K^{\prime}$, where they mix with the 0th pLLs. This can be understood as follows. In the continuum approach, for a real magnetic field the guiding center of the LLs is $y_0(p_x,\tau) =  p_x \ell^2$, where $\ell$ is the magnetic  length. Thus, independently of the valley from which they originate, the states with $p_x > 0$ (i.e. $y_0 > 0$) and $p_x < 0$ (i.e. $y_0 < 0$) are localized on opposite sides relative to the center of the ribbon (which corresponds to $y_0 = 0$ in the continuum limit). As $k_x$ shifts away from the center of the BZ, the two edge-state bands at opposite edges evolve into the two 0th LLs at opposite valleys, which are degenerate because of inversion symmetry (cf. Fig.~\ref{fig:fig3}). On the other hand, for nonuniform strain, the guiding center is $y_0(p_x,\tau) =  \tau p_x \ell^2$ (cf. Eq.~\eqref{eq:opy}). Hence, going from one valley to the other (i.e. $\tau \to -\tau$) is akin to $p_x \to -p_x$, and therefore $y_0(p_x, -\tau) = y_0(-p_x,\tau) = -y_{0}(\tau, p_x)$. That is, from one valley to the other, the guiding center changes sign relative to the center of the ribbon. This requires that the same band of edge states (here, the top edge one)  evolves into the 0th pLL at both valleys and thus crosses the gap connecting them. However, the other edge band does not mix with the pLLs and does not cross them either. In this regard,  our observations differ from the conclusion reached by Ghaemi et al.,~\cite{Ghaemi13} who solved the Dirac equation and argued that the edge-state bands should be connected to high energy states and cross
the 0th pLL without hybridizing with it. As shown in the following section, such description is also not accurate when applied to the edge bands for the armchair ribbon.
  
% Qualitatively, the difference in the edge band dispersion can be also understood as follows. The top edge band belongs to states localized at the edge that is maximally stretched and where the hopping parameters are locally reduced from their bulk values, $t$ and $t^{\prime}$ (cf. Eqs.~\ref{eq:st} and \ref{eq:stp}). 
%\textcolor{red}{Since the local hopping determines the group velocity of the edge states, the decrease in the local hoping leads to a (slight) decrease group velocity.  On the other hand, the bottom  edge is maximally compressed, which effectively increases the hopping parameters and thus the group velocity.}
%\textcolor{blue}{The top edge band distributes over a narrower range of $p_x$ since the lattice is maximally expended. The bottom edge band, on the other hand, distributes over a wider range of $p_x$ since the lattice is maximally compressed.}
%Notice that for a magnetic field (cf. Fig.~\ref{fig:fig3}(b)) this asymmetry does not exist. 

%

\begin{figure}
\centering
\includegraphics[scale=0.48]{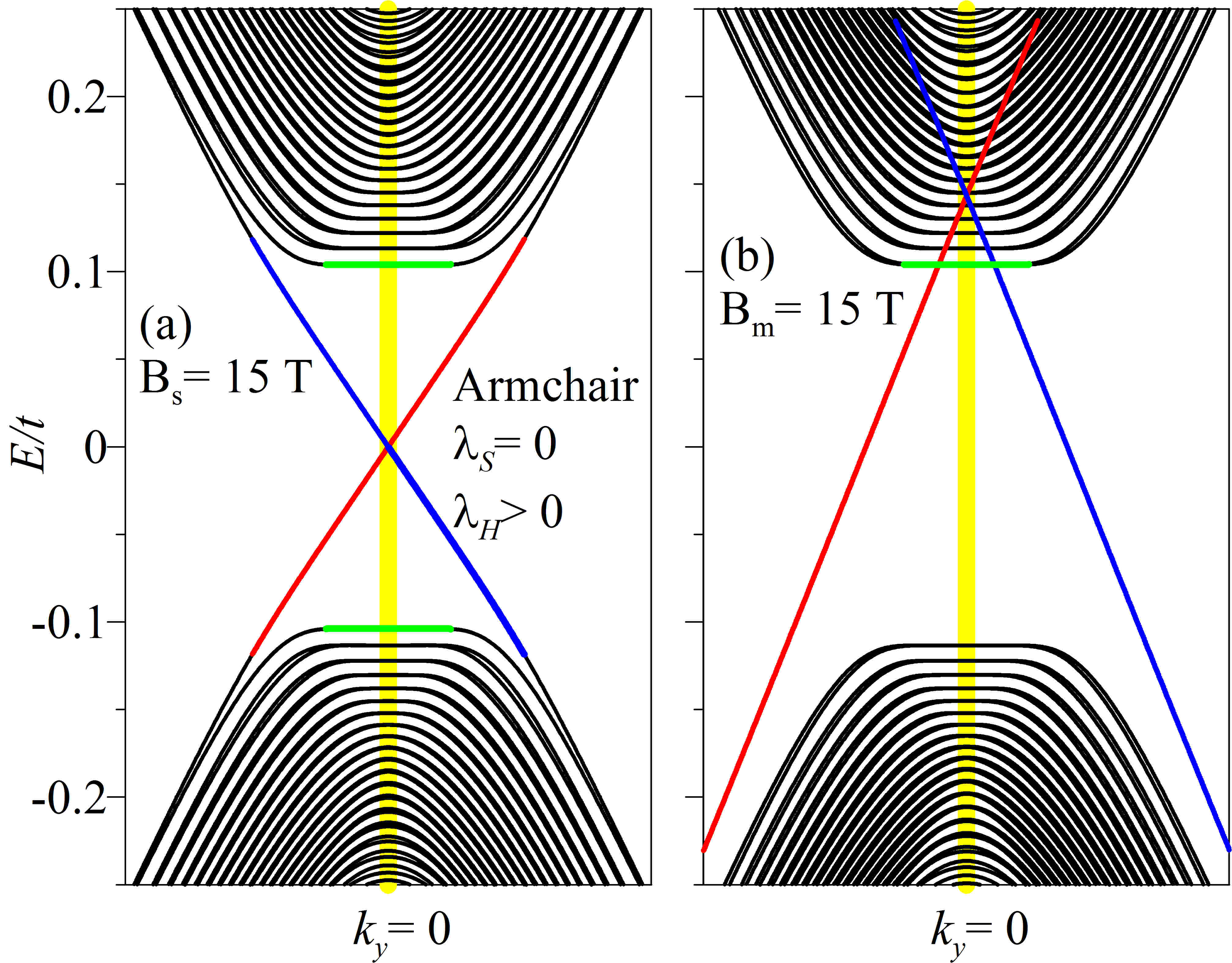}
\caption{(Color online) Band structure of an armchair ribbon described by the Haldane model subject to nonuniform shear strain. Model parameters are the same as in Fig.~\ref{fig:fig3}.}
\label{fig:fig4}
\end{figure}
 
\begin{figure}
\centering
\includegraphics[scale=0.48]{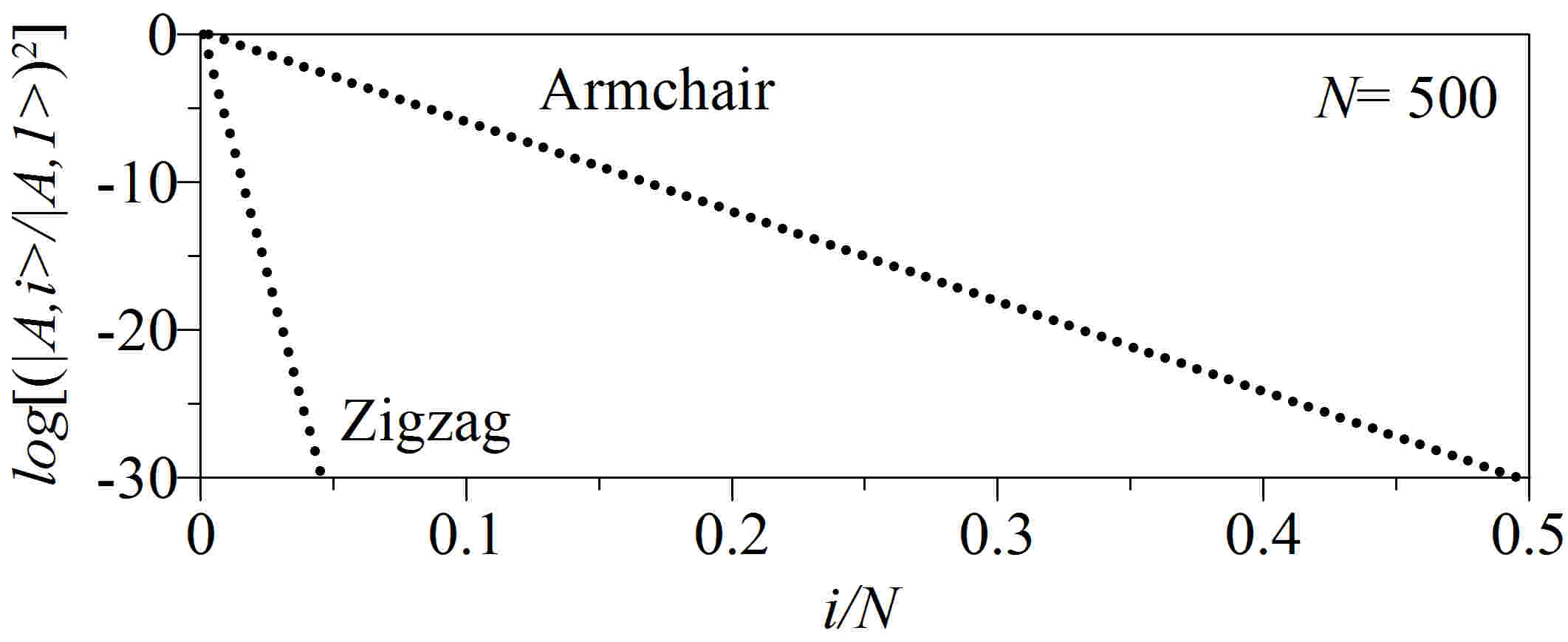}
\caption{(Color online) Normalized spatial wavefunction (on sublattice $A$) of edge states, showing that the penetration from the topological edges into the bulk is very different for the zigzag and armchair edges. The edge state taken is a quarter way between $K$ and $K'$ for zigzag case and $k_y$=0 for armchair case.}
\label{fig:fig5}
\end{figure}

\subsection{\bf Armchair Ribbon}

 Along the armchair direction, the $K$ and $K^{\prime}$ points are projected on $k_y = 0$, i.e. at center of the BZ. In the continuum limit, the armchair boundary conditions couple the $K$ and $K'$ valleys to each other.~\cite{Neto09} Therefore, unlike the zigzag ribbon, there is only one conduction band minimum and one valence band maximum at $k_y = 0$. The gap between the two bands is controlled by the Haldane mass. Inside the gap, a pair of topologically protected \emph{edge}-state bands, corresponding to the edge states at the left and right edge, intersect at $k_y = 0$. 
 
  The band structure of a strained armchair ribbon described by the HM is shown in Fig.~\ref{fig:fig4}. Application of nonuniform shear strain in this case also leads to the appearance of pLLs, see Fig.~\ref{fig:fig4}(a). Unlike the case of a real magnetic field (cf. Fig.~\ref{fig:fig4}(b)), the 0th pLLs split from the bottom  and top of the conduction and valence band, respectively, and are therefore located symmetrically relative to the middle of the gap, in agreement with the predictions of the continuum model (see Appendix~\ref{app:0ll}). However, there is no tilt of the pLLs. This is because for the armchair direction strain is of the shear type (i.e. traceless), and therefore the deformation potentials of Eq.~\eqref{eq:sgap} vanish. In addition, the boundary conditions mix the two valleys and thus the analysis becomes more involved than the one given in Appendix~\ref{app:tilt}, which assumes the valley index $\tau$ is a good quantum number.

 As for the armchair ribbon, we find that strain does not strongly modify dispersion of the edge-state bands. However, unlike the zigzag ribbon, in this case none of
 edge-state connects to the 0th pLLs (and they also  do not cross the 0th LLs without hybridizing as suggested in Ref.~\onlinecite{Ghaemi13}).  Fig.~\ref{fig:fig4}(a) shows that the effects of shear strain are also qualitatively different as the energy of the in-gap states is not shifted near $k_y =0$,  as it happens when the ribbon is subject to a real magnetic field, see Fig.~\ref{fig:fig4}(b) (however, notice that in this case the edge-state bands are connected to high energy states and  do cross the 0th LLs without hybridizing with it). 
 
  It is also interesting to notice that in the armchair case the penetration of the edge states into the bulk is much larger than for the zigzag edge (cf. Fig.~\ref{fig:fig5}). Indeed, for the armchair edge, the penetration is determined by the band gap (i.e. $t^{\prime}$), whereas for the zigzag edge, the edge states are already present for $t^{\prime} = 0$, and the penetration is determined by the bandwidth ($\sim t$). Nevertheless, despite of longer penetration, nonuniform strain has a very weak effect on the dispersion. 

\section{Effects of strain at criticality}\label{sec:tt}

 Finally, in this section, we shall study the effects of nonuniform strain on the critical point of the topological phase transition, i.e. for $\lambda_S = \lambda_H$. The Semenoff mass breaks (sublattice) inversion symmetry and lifts the degeneracy between the two valleys (cf. Fig.~\ref{fig:fig1}(d)). At the critical point, the gap closes at one of the valleys (cf. Fig.~\ref{fig:fig1}(d)) and the system becomes a metal. 
 
\begin{figure}
\centering
\includegraphics[scale=0.48]{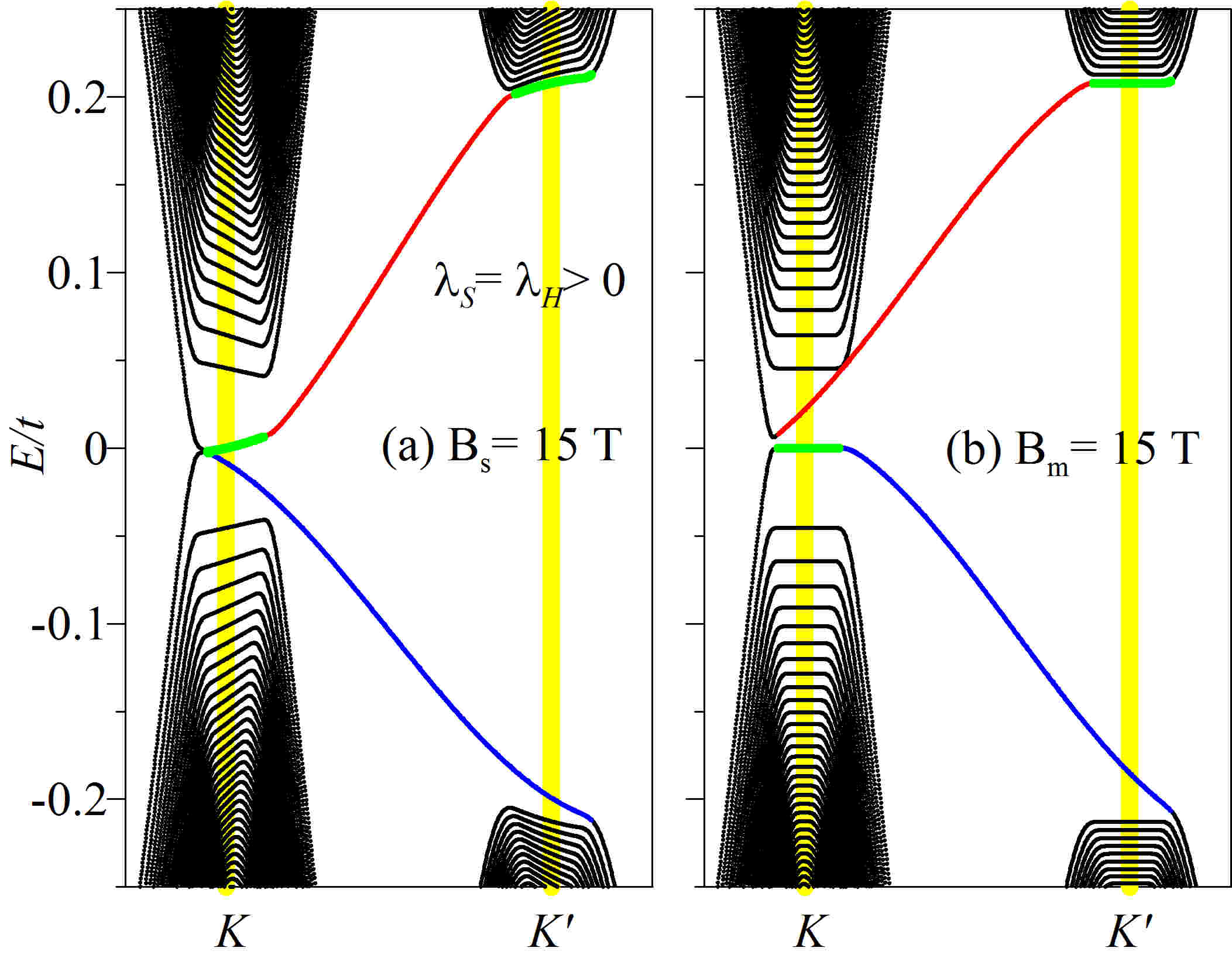}
\caption{(Color online) Effect of nonuniform strain on the band structure of a zigzag ribbon at the topological critical point. (a) Under nonuniform strain leading to a pseudo-magnetic field, pseudo Landau levels emerge, with the edge  bands becoming degenerate near the $K$ point. (b) Under a real magnetic field, the degeneracy of the edge bands is lifted. Parameters are the same as for Fig.~\ref{fig:fig3}. However, the value of the inversion symmetry breaking potential is tuned to the critical point of the topological transition where $\lambda_S=\lambda_H$ (cf. Eq.~\eqref{eq:kpham}).}
\label{fig:fig6}
\end{figure}

 Nevertheless, even though the band gap closes in one of the valleys (here $K$), the band structure near $K$ is still very different from strained graphene (cf. Fig.~\ref{fig:fig2}(a)). For the strained zigzag edge, Fig.~\ref{fig:fig6}(a) shows that the 0th pLL is tilted. The origin of this tilt is again the shift $\Delta \epsilon^{\prime}_{n\pm}(\tau = +1)$ discussed in Sec.~\ref{sec:strainhm} (cf. Eq.~\eqref{eq:tiltp}), which originates from the deformation potential that corrects the (topological) band gap (i.e. the term proportional to $\gamma^{\prime\prime} \tau \sigma_z$ in Eq.~\eqref{eq:sgap}). Note that, although in the present tight-binding calculation the other terms ($\gamma$ and $\gamma^{\prime}$) are set to zero, this result will not change for non-vanishing $\gamma$ and $\gamma^{\prime}$ unless fine tuning of these deformation potentials nullifies the correction arising from $\mathcal{H}^{\prime}$ in Eq.~\eqref{eq:sgap}. 

\begin{figure}
\centering
\includegraphics[scale=0.32]{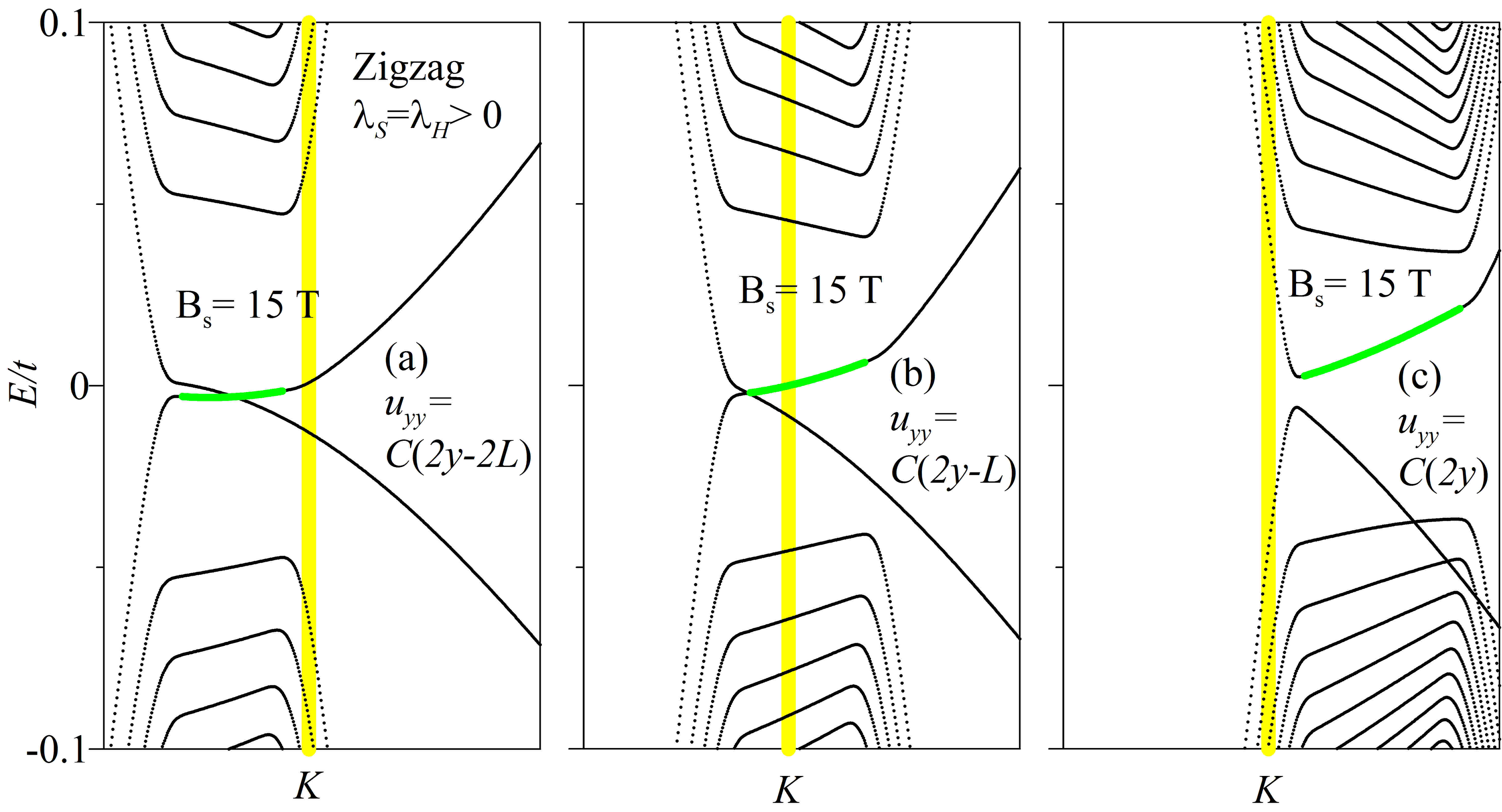}
\caption{(Color online) Band structure of a zigzag ribbon described by the Haldane model at the topological phase boundary (i.e. for $\lambda_S=\lambda_H=3\sqrt{3}t'$) subject to a pseudo magnetic field $\mathcal{B}_s$=15 T originating from nonuniform uniaxial strain generated by (a) $u_{yy}=2C(y-L)$ (b) $u_{yy}=2C(y-{L\over 2})$ and (c) $u_{yy}=2C \: y$.}
\label{fig:fig7}
\end{figure}

 In Fig.~\ref{fig:fig6}, we also compare the effects of a real magnetic field and nonuniform strain for the zigzag edge. Notice that whereas the magnetic field splits the degeneracy between the edges state bands at opposite sides of the ribbon and opens a (small) gap, strain does not. Thus, it seems that the system remains gapless after the application of nonuniform strain. In order to further investigate whether the band touching is robust against the type of strain configuration, we consider three different strain configurations corresponding to different parts of the ribbon being clamped and strained, respectively shown in Fig.~\ref{fig:fig7}. The strain configuration used in Fig.~\ref{fig:fig5}(a) and Fig.~\ref{fig:fig6}(b), i.e. $u_{yy} = C(2y-L)$, corresponds to the center of the ribbon being clamped. while one of the edges is stretched and the other compressed. In this case, the bottom-edge band touches the 0th pLL in the neighborhood of the $K$ point and the system remains gapless. However, for the strain configuration corresponding to $u_{yy} = C(2y- 2L)$, for which the upper edge at $y = L$ is clamped and the bottom edge is compressed, the bottom-edge band crosses the 0th pLL. Finally, for $u_{yy} = 2 C y $ (bottom edge clamped and lower edge stretched), a small gap opens in the neighborhood of $K$ and the behavior of the lowest subbands under strain is reminiscent of the case of a real magnetic field. Thus, we conclude the band touching is not robust and actually depends on the strain configuration applied to the ribbon.


\section{Conclusions}~\label{sec:concl}

 Nonuniform strain induces a pseudo-magnetic field and pseudo-Landau levels in the band structure of strained ribbons described by the Haldane model. However, unlike a real magnetic field, the band structure of ribbons under strain shows distinct features arising from a breaking of two-dimensional inversion symmetry. One such feature is the absence of degenerate zeroth (pseudo-) Landau levels, which is characteristic of a real magnetic field~\cite{Haldane88}. However, unlike the inversion-symmetry breaking induced by a staggered potential (i.e. a ``Semenoff mass''), the magnitude of the band gap appears to be weakly modified and strain does not strongly affect the in-gap edge states. This indicates that the topologically non-trivial phase is robust against strain, thus confirming an observation  made earlier  Ghaemi and coworkers,~\cite{Ghaemi13} who relied on the continuum-limit Dirac equation. 
 
 However, departing from the conclusions reached in Ref.~\onlinecite{Ghaemi13}, we find a very different interplay of the edge state bands with the pseudo-Landau levels. 
For instance, our tight-binding results show that the topologically protected edge-state bands in both zigzag and armchair ribbons do not cross the zeroth pseudo-Landau level without hybridizing with it. Instead, for the zigzag ribbon one of the branches of edge-state smoothly evolves into  bulk states by connecting the two zeroth pseudo-Landau levels, whereas the other band remains entirely localized at the edge. We also observe that the pseudo-Landau levels are tilted along the zigzag direction, which lifts the degeneracy of the bulk states with the pseudo-Landau level. The dominant contribution to this tilt for the lowest pseudo-Landau levels arises from the deformation potential that also modifies the size of the gap under uniaxial strain. For higher Landau levels, the strain-induced correction to the Fermi velocity which accounts for the tilt in graphene becomes dominant. 
For the armchair ribbon, the dispersion of the in-gap states is barely modified from their unstrained form, despite their large penetration into the bulk.  In this case, as for the zigzag ribbon,  we also find the edge-bands do not cross the zero-th pseudo-Landau level, unlike the case of a real magnetic field.  

Finally, at the critical point of the topological phase transition where one of the valleys becomes gapless, we have shown that nonuniform strain has different effects on the pseudo-Landau levels than for the purely gapless graphene. Some of the differences, like the tilt of the zeroth pseudo-Landau level, can be traced back to the deformation potentials that are present in the Haldane model but not in graphene.

\acknowledgments 
We thank Maria A. H. Vozmediano for fruitful discussions.
Y.-H.H. and M.A.C. acknowledge the support from the Ministry of Science and Technology (Taiwan) under contract number NSC 102- 2112-M-007-024-MY5, and Taiwan's National Center of Theoretical Sciences (NCTS). E.V.C. acknowledges partial support from
FCT-Portugal through Grant No. UID/CTM/04540/2013.

\appendix 
\renewcommand\thefigure{\thesection.\arabic{figure}}

\section{Details}

\subsection{Band gap and Fermi velocity corrections}\label{app:gap} \label{app:modgap}
\setcounter{figure}{0}

%\begin{figure}
%\centering
%\includegraphics[scale=0.32]{M5FS2}
%\caption{(Color online) Haldane subbands near the topological phase boundary ($%\lambda_S=\lambda_H=3\sqrt{3}t'$) with (a) a magnetic field $\mathcal{B}_m$=15 T %and a pseudo magnetic field $\mathcal{B}_s$=15 T of the strain type (b) $u_{yy}%=2C(y-L)$ (c) $u_{yy}=2C(y-{L\over 2})$ and (d) $u_{yy}=2C(y)$.}
%\label{fig:fig4}
%\end{figure}

Within the $\vec{k}\cdot\vec{p}$ approach, in the absence of strain, the next nearest-neighbor hopping term (cf. Eq.~\eqref{eq:hm3}) yields the following contribution to the continuum Hamiltonian:
\begin{align}
\mathcal{H}_{g}&=3t'(\cos\psi\sigma_0+\sqrt{3}\sin\psi\tau\sigma_z)
\end{align}
The term $\mathcal{H}_{g}$ gives rise to the Haldane mass, which gives rise to the band gap $E_g=6\sqrt{3}t'\sin\psi$. The contribution due to the strain arising from the second neighbor hoping read: 
\begin{align}
\mathcal{H}^{\prime} &={3\over 2}\beta't'(u_{xx}+u_{yy})(\cos\psi\sigma_0+\sqrt{3}\sin\psi\tau\sigma_z),\\
\mathcal{H}^{\prime\prime} &={{3\sqrt{3}}\over 4}a\beta't' \left[-(u_{xx}-u_{yy})p_x+2u_{xy}p_y\right]\notag \\
&\quad \times (\sqrt{3}\cos\psi\tau\sigma_0-\sin\psi\sigma_z),
\end{align}
where we have taken $\beta^{\prime} = \tfrac{d\log t^{\prime}}{d \log\sqrt{3}a}\simeq -1$ as the Gr\"uneisen parameter for the next nearest-neighbor hoping ($t^{\prime}$). $\mathcal{H}^{\prime}$ yields correction of the size of the band gap for the zigzag ribbon. However, for the armchair ribbon with $\psi=\pi/2$ and $u_{xy}\ne 0$, $\mathcal{H}^{\prime} =0$ and therefore the band gap of the strained armchair is not affected by strain. Furthermore, $\mathcal{H}^{\prime\prime}$ yields a correction to the Dirac cone tilt, and thus the Fermi velocity.

\subsection{$0$th Landau level in strained Haldane}
\label{app:0ll}

In the presence of strain, the Haldane model in the continuum limit takes the form:
\begin{equation}
\mathcal{H} = v \left( \tau\sigma_x \Pi_x + \sigma_y \Pi_y \right) - \lambda_H \sigma_z\tau + \lambda_S \sigma_z, 
\end{equation}
where ${\bf \Pi}= {\bf p} + g\vec{\mathcal{A}}$ and $\vec{\mathcal{A}}$ the pseudo-gauge field. In terms of the strain tensor
\begin{equation}
\vec{\mathcal{A}} \propto \tau\left( - u_{xx}+ u_{yy}, 2u_{xy}\right).  
\end{equation}
For the zigzag direction, $u_y\propto y^2$, this leads to a pseudo-magnetic field:
\begin{equation}
\vec{\mathcal{B}}_s=\nabla\times\vec{\mathcal{A}}=-\mathcal{B}_s\tau\hat{z}.
\end{equation}
For the above choice of the atomic displacement field, the problem of diagonalizing $\mathcal{H}$ in each valley is akin to solving the 2D Dirac equation for the (pseudo) Landau level (pLL) in the Landau gauge, i.e.
\begin{equation}
\vec{\mathcal{A}}= \tau (\mathcal{B}_s y,0).
\end{equation}
Thus, the Hamiltonian takes the matrix form ($g = \frac{2}{3})$:
\begin{equation}
\mathcal{H} = \left(
\begin{array}{cc}
-\lambda_H\tau + \lambda_S & v \left[\tau (p_x+g \mathcal{B}_s\tau y)-ip_y\right] \\
v \left[\tau (p_x+g \mathcal{B}_s\tau y)+ip_y\right] & \lambda_H\tau - \lambda_S
\end{array}
\right).
\end{equation}
Note that, in the Landau gauge, $p_x$ is a constant of motion, i.e. $\left[p_x,\mathcal{H}\right] = 0$. The operators 
\begin{equation}
\pi^{\pm} = \frac{1}{\sqrt{2g|\mathcal{B}_s|}}\left( \tau p_x+g \mathcal{B}_s y \pm ip_y \right),
\end{equation}
obey
\begin{equation}
\left[\pi^{+},\pi^{-} \right] 
=  \mathrm{sgn}(\mathcal{B}_s),
\end{equation}
Thus, the sign of $\mathcal{B}_s$ determines which operator of the pair $\Pi^{\pm}$ behaves as the raising and lowering operator; e.g. for $\mathcal{B}_s > 0$, $\pi^{+}$ ($\pi^{-}$) behaves as the pLL lowering (raising) operator. It is convenient to introduce the cyclotron frequency $\omega_c = v \sqrt{2g | \mathcal{B}_s|}$ and $\Delta_{\tau} = \lambda_H \tau - \lambda_S$, which allows us to write:
\begin{equation}
\mathcal{H} =\left(
\begin{array}{cc}
-\Delta_{\tau} & \omega_c \pi^{-}\\
\omega_c \pi^{+} & \Delta_{\tau}
\end{array}
\right).
\end{equation}
Hence,
\begin{align}
\mathcal{H}^2 = \left(
\begin{array}{cc}
 \Delta_{\tau}^2 + \omega^2_c \pi^{-}\pi^{+} & 0\\
 0 & \Delta^2_{\tau} + \omega^2_c \pi^{+} \pi^{-}
\end{array}
\right). 
\end{align}
Let us consider the two possible cases, i.e. $\mathcal{B}_s < 0$ and $\mathcal{B}_s > 0$. For $\mathcal{B}_s > 0$, we can identify $a = \pi^{+}$ (i.e. lowering) and $a^{\dag} = \pi^{-}$ (i.e. raising). Hence,
\begin{align}
\mathcal{H} &=  \left( 
\begin{array}{cc}
-\Delta_{\tau} &  \omega_c a^{\dag}\\
\omega_c a & \Delta_\tau  
\end{array}
\right),\\
\mathcal{H}^2 &=  \left(
\begin{array}{cc}
\Delta_{\tau}^2 + \omega^2_c n & 0 \\
0 & \Delta^2_{\tau} + \omega^2_c \left( n + 1 \right)
\end{array}
\right).
\end{align}
Hence, the eigenvectors are:
\begin{equation}
|\Psi_{n\pm}(\tau)  \rangle = \frac{1}{\sqrt{2(1+\lambda^2_{n\pm}(\tau)})}\left( \begin{array}{c}
|n \rangle \\ \lambda_{n\pm}(\tau) |n-1\rangle 
\end{array}
\right) \label{eq:pllstates}
\end{equation}
where 
\begin{equation}
\lambda_{n\pm}(\tau) =\frac{\Delta_{\tau} \pm \sqrt{\Delta_{\tau}^2 +\omega^2_c n}}{\omega_c \sqrt{n}},
\label{eq:lambda}
\end{equation}
and  energy eigenvalue $\epsilon_n = \pm \omega_c \sqrt{ \Delta_{\tau}^2 + \omega^2_c n}$ for $n\neq 0$. However, the 0th pLL requires a separate treatment. In such case, the following state
\begin{equation}
|\Psi_{0}(\tau) \rangle = \left( \begin{array}{c}
|0 \rangle \\0 
\end{array}
\right), 
\end{equation}
which satisfies
\begin{align}
\mathcal{H}  |\Psi_{0}(\tau) \rangle = - \Delta_{\tau} |\Psi_{0}(\tau)\rangle 
\end{align}
is an eigenstate with energy $\varepsilon_0 = -\Delta_{\tau} = -\lambda_H \tau + \lambda_S$. On the other hand, for $\mathcal{B}_s < 0$, we have $a = \pi^{-}$ (i.e. lowering) and $a^{\dag} = \pi^{+}$ (i.e. raising), which leads to
\begin{align}
\mathcal{H} &=  \left( 
\begin{array}{cc}
-\Delta_{\tau} &  \omega_c a\\
\omega_c a^{\dag} & \Delta_{\tau}
\end{array}
\right),\\
\mathcal{H}^2 &=  \left(
\begin{array}{cc}
\Delta_{\tau}^2 + \omega^2_c \left( n +1\right) & 0 \\
0 & \Delta_{\tau}^2 + \omega^2_c n 
\end{array}
\right),
\end{align}
Focusing again on the 0th pLL, which is described by the state:
\begin{equation}
|\Psi_{0}(\tau) \rangle = \left( \begin{array}{c}
0\\ |0 \rangle  
\end{array}
\right) 
\end{equation}
we have
\begin{align}
\mathcal{H} |\Psi_{0}(\tau) \rangle =  \Delta_{\tau} |\Psi_{0}(\tau)\rangle.
\end{align}
and hence $\varepsilon_0 = \lambda_H\tau- \lambda_S$. The above results can be summarized in the following expression:
\begin{equation}
\varepsilon_0 = -\Delta_{\tau} \mathrm{sgn}(\mathcal{B}_s) = \left( \lambda_S - \lambda_H \tau \right)  \mathrm{sgn}(\mathcal{B}_s),
\end{equation}
which implies that, for given $\mathrm{sgn}(\mathcal{B}_s)$ and $\lambda_S = 0$, the 0th pLLs have opposite energy (relative to the middle of the gap) in opposite valleys. It is also worth noticing that the wavefunctions of pLLs in both valleys have the same sublattice structure.

 On the other hand, repeating the calculation for real magnetic field in the Landau gauge field, i.e.
\begin{equation}
\vec{\mathcal{A}}_m = \left(-\mathcal{B}_m y, 0\right),
\end{equation}
yields (for $\lambda_S = 0$)
\begin{equation}
\varepsilon_0 =  \lambda_H \, \mathrm{sgn}(\mathcal{B}_m),
\end{equation}
which implies that the 0th LLs in different valleys are degenerate in energy. In addition, the wavefunctions of the LLs on different valleys have the opposite sublattice structure, which is different from the case of strain.

\subsection{Tilt of pseudo-Landau levels}\label{app:tilt}
\setcounter{figure}{0}
 
 In order to explain the tilt exhibited by the pLLs in the zigzag edge, which is not present for the LLs (cf. Fig. 2(b), for the zigzag edge), we need to take into account the lowest order treatment of the strain, introduced in Sec.~\ref{sec:strainhm}. For the zigzag edge under uniaxial strain (i.e. $u_{xx}=u_{xy}=0$ and $u_{yy} = 2C y \neq 0$), Eq.~\eqref{eq:strain_veloc} reduces to
\begin{align}
 \mathcal{H}^{\prime\prime} &={3\over 8 } \left( \beta t\right) u_{yy} \left( \tau \sigma_x p_x a   + 3 \sigma_y p_ya \right),
\end{align}
Thus, to leading order in perturbation theory, the energy shift in the pLL energy is:
\begin{align}
\Delta \epsilon^{\prime\prime}_{n\pm} (\tau) &= \langle \Psi_{n\pm}(\tau) |  \mathcal{H}^{\prime\prime} |\Psi_{n\pm}(\tau)\rangle \\
& = -\frac{3}{2}  C \ell  \beta t    \frac{ \tau  \lambda_{n\pm}(\tau)  \sqrt{n} }{1+\lambda^2_{n\pm}(\tau)}\:   \left( p_x a \right),  \label{eq:corra}
\end{align}
where $\ell = \left(2 g \mathcal{B}_s\right)^{-1/2}$ is the (pseudo-) magnetic length. In Eq. (A9), we have used (see Appendix~\ref{app:0ll}) that (recall that $p_x$ is a constant of motion):
\begin{align} 
y &=   \ell \left(a + a^{\dag} \right)  - 2 \tau p_x
\ell^2   \label{eq:opy}, \\
p_y &=  \frac{\left(a - a^{\dag} \right)}{2i\ell}. 
\end{align}
In the above expressions, $a$ ($a^{\dag}$) is the lowering (raising) operator for pLLs. Note that the correction in Eq.~\eqref{eq:corra} is linear in $p_x$ and proportional to $\tau$. Since the expression of $\lambda_{n\pm}(\tau)$ diverges for $n\to 0$, the 0th pLLs require a separate treatment, and thus
\begin{equation}
\Delta \epsilon^{\prime\prime}_0(\tau) = 0.
\end{equation}
which could have be obtained from Eq.~\eqref{eq:corra} by taking $\lambda_{n\pm}(\tau) \to 0$ for $n \to 0$.

In addition, since $y$ contains a term proportional to $p_x$ (cf. Eq.~\eqref{eq:opy}), there is an linear in $p_x$ contribution to the energy shift of the pLLs arising from the deformation potentials in Eq.~\eqref{eq:sgap}. For the zigzag HM ribbon with $\psi = \tfrac{n\pi}{2}$ and $\epsilon_0 = 0$, we show in the following Appendix that the deformation potential takes the form:
\begin{equation}
 \mathcal{H}^{\prime} = 
\frac{3\sqrt{3}}{2} \beta^{\prime} t^{\prime} \tau\sigma_z u_{yy} 
\end{equation}
To linear order in $p_x$, the energy shift resulting from this term reads:
\begin{align}
\Delta \epsilon^{\prime}_{n\pm}(\tau) &= \langle \Psi_{n\pm} (\tau)|  \mathcal{H}^{\prime} |\Psi_{n\pm}(\tau)\rangle \\
 &=-3\sqrt{3} C\ell  \left( \beta^{\prime} t^{\prime}\right) \left( \frac{1-\lambda^2_{n}(\tau)}{1+
 \lambda^2_{n}(\tau)}\right)  \left(p_x \ell\right), \label{eq:corrb}
\end{align}
where $\lambda_{n\pm}(\tau)$ is given in Eq.~\eqref{eq:lambda}. It can be seen that $\Delta \epsilon^{\prime}_{n\pm}(\tau) = \Delta \epsilon^{\prime}_{n\mp}(\tau)$. However, note  the 0th pLL energy correction $\Delta \epsilon_{n = 0}(\tau)$, which can be also obtained by setting $\lambda_{n\pm}(\tau) = 0$ in Eq.~\eqref{eq:corrb}, remains finite.  

In order to prove the other relations given in Eqs.~(\ref{eq:rel2},\ref{eq:rel3}) we first observe that, for $\lambda_{S} = 0$, 
\begin{equation}
\lambda_{n\pm}(\tau) = -\lambda_{n\mp}(-\tau).
\end{equation}
Therefore, $\tau \lambda_{n\pm}(\tau) = \tau  \lambda_{n\pm}(-\tau)$. This  allows to prove the second equation, i.e. Eq.~\eqref{eq:rel2}.

Finally, we notice that, strictly speaking, the above solution within the Landau gauge only applies to the bulk states as we have not taken into account the modifications to the Landau 
level dispersion arising from the  boundary conditions. The latter are important for pLLs localized near boundary and explain the dispersion of the pLLs far from the $K$ and $K^{\prime}$ points.


\begin{thebibliography}{99}
%
\bibitem{Haldane88}  F. D. M. Haldane, Phys. Rev. Lett. ${\bf 61}$, 2015 (1988).
%
\bibitem{KaneMele}
C.~L. Kane and E.~J. Mele, Phys. Rev. Lett. {\bf 95}, 226801 (2005).
%
\bibitem{Bernevig}
B.~A. Bernevig with T.~L Hughes, \emph{Topological Insulators and Topological Superconductors}, Princeton University Press (Princeton, 2013).
%
\bibitem{Hasan2010rmp} M. Z. Hasan and C. L. Kane, Rev. Mod. Phys. \textbf{82%
}, 3045 (2010); X.-L. Qi and S.-C. Zhang, Rev. Mod. Phys. \textbf{83}, 1057
(2011).
%
\bibitem{Konig}
M. K\"onig, H. Buhmann, L.~ W. Molenkamp, T. Hughes, C.-X. Liu, 
X.-L. Qi, and S.~C. Zhang, J. Phys. Soc. Jpn. {\bf 77}, 031007 (2008).
M. K\"onig \textit{et al.}, Science \textbf{318}, 766
(2007); A. Roth \textit{et al.} Science \textbf{325}, 294
(2009); K. C. Nowack \textit{et al.} Nat.~Mater.  {\bf 12}, 787 (2013); G.Grabecki \textit{et al.}, Phys. Rev. B {\bf 88}, 165309 (2013);  G. M. Gusev \textit{et al.} Phys. Rev. B {\bf 89}, 125305 (2014); I. Knez, R.-R. Du, and G. Sullivan, Phys. Rev. Lett. {\bf 107}, 136603 (2011); K. Suzuki, Y. Harada, K. Onomitsu, and K. Muraki,
Phys. Rev. B {\bf 87}, 235311 (2013); I. Knez \textit{et al.} Phys. Rev. Lett. {\bf 112}, 026602 (2014); E. M. Spanton \textit{et al.} Phys. Rev. Lett. {\bf 113}, 026804
(2014).
%
\bibitem{Weeks}
C. Weeks, J. Hu, J. Alicea, M. Franz, and R. Wu,  Phys.~Rev. X {\bf 1}, 021001 (2011).
%
\bibitem{Cresti}
A. Cresti, D.~V. Tuan, D. Soriano, A.~W. Cummings, S. Roche Phys. Rev. Lett. {\bf 113}, 246603 (2014).
%
\bibitem{Brey}
L. Brey, Phys.Rev.B {\bf 92}, 235444 (2015).
%
\bibitem{Garcia}
J. H. Garcia and T.~G. Rappoport, 2D Materials {\bf 3}, 024007 (2016).
%
\bibitem{Jotzu14}    G. Jotzu, M. Messer, R. Desbuquois, M. Lebrat, T. Uehlinger, D. Greif, and T. Esslinger, Nature ${\bf 515}$, 237 (2014).
%
\bibitem{Semenoff84} G. V. Semenoff, Phys. Rev. Lett. ${\bf 53}$, 2449 (1984).
%
\bibitem{Zhai10}     F. Zhai, X. Zhao, K. Chang, H.-Q. Xu, Phys. Rev. B ${\bf 82}$, 115442 (2010).
%
\bibitem{Chaves10}   A. Chaves, L. Covaci, Kh. Yu. Rakhimov, G. A. Farias, and F. M. Peeters, Phys. Rev. B ${\bf 82}$, 205430 (2010).
%
\bibitem{Guinea10}   F. Guinea, M. I. Katsnelson, A. K. Geim, Nature Phys.{\bf 6} 30 (2010).
%
\bibitem{Suzuura02}  H. Suzuura, T. Ando, Phys. Rev. B ${\bf 65}$, 235412 (2002).
%
\bibitem{Low10}      T. Low, and F. Guinea, Nano Lett. ${\bf 10}$, 3551 (2010).
%
\bibitem{Ghaemi13}   P. Ghaemi, S. Gopalakrishnan, and S. Ryu, Phys. Rev. B ${\bf 87}$, 155422 (2013).
%
\bibitem{xianpeng17} X.-P. Zhang,  H. Chunli, and M. A. Cazalilla,  2D Materials, {\bf 4} 024007 (2017).
%
\bibitem{castro17}   E.~V. Castro, M. A. Cazalilla, and M.~A.~H. Vozmediano, report arxiv:1610.08988 (2016).
%
\bibitem{cog_tmdc}   M. A. Cazalilla, H. Ochoa, and F. Guinea, Phys.~Rev.~Lett. {\bf 113}, 077201 (2014).
%
\bibitem{Castro16}   L. Li, E. V. Castro, and P. D. Sacramento, Phys. Rev. B ${\bf 94}$, 195419 (2016).

\bibitem{Amorim16}   B. Amorim, A. Cortijo, F. de Juan, A. G. Grushin, F. Guinea, A. Gutierrez-Rubio, H. Ochoa, V. Parente, R. Roldan, P. San-Jose, J. Schiefele, M. Sturla, M. A. H. Vozmediano, Phys. Rep. ${\bf 617}$, 1 (2016).
%
\bibitem{Ho11}       Y.-H. Ho, J. Wang, Y.-H. Chiu, M.-F. Lin, and W.-P. Su, Phys. Rev. B ${\bf 83}$, 121201 (2011); Y.-H. Ho, ${\it ibid}$. ${\bf 87}$, 075417 (2013); Y.-H. Ho, J.-Y. Wu, R.-B. Chen, Y.-H. Chiu, and M.-F. Lin, Appl. Phys. Lett. ${\bf 97}$, 101905 (2010); Y.-H. Ho, ${\it ibid}$. ${\bf 99}$, 011914 (2011); Y.-H. Ho, J.-Y. Wu, Y.-H. Chiu, J. Wang, and M.-F. Lin, Philos. Trans. R. Soc. A ${\bf 368}$, 5445 (2010).
%
\bibitem{Liu15}     R.-B. Liu, D.-L. Deng, D.-W. Zhang, and S.-L. Zhu, J. Opt. Soc. Am. B ${\bf 32}$, 2500 (2015).
%
\bibitem{Liu11}     C.-C. Liu, W. Feng, and Y. Yao, Phys. Rev. Lett. ${\bf 107}$, 076802 (2011) ; C.-C. Liu, H. Jiang, and Y. Yao, Phys. Rev. B ${\bf 84}$, 195430 (2011).
%
\bibitem{Tabert13}   C. J. Tabert and E. J. Nicol, Phys. Rev. Lett. ${\bf 110}$, 197402 (2013); L. Stille, C. J. Tabert, E. J. Nicol, Phys. Rev. B ${\bf 86}$, 195405 (2012).
%
\bibitem{Ezawa12}    M. Ezawa, New. J. Phys. ${\bf 14}$, 033003 (2012).
%
\bibitem{Barkhofen13}S. Barkhofen, M. Bellec, U. Kuhl, and F. Mortessagne, Phys. Rev. B ${\bf 87}$, 035101 (2013).
%
\bibitem{Goerbig11}  M. O. Goerbig, Rev. Mod. Phys. ${\bf 83}$, 1193 (2011).
%
\bibitem{Neto09}     A. H. Castro Neto, F. Guinea, N. M. R. Peres, K. S. Novoselov and A. K. Geim, Rev. Mod. Phys. ${\bf 81}$, 109 (2009).
%
%\bibitem{deJuan13}   F. de Juan, J. L. Manes, M. A. H. Vozmediano, Phys. Rev. B ${\bf 87}$, 165131 (2013).
\bibitem{deJuan13}   J. L. Manes, F. de Juan, M. Sturla, M. A. H. Vozmediano, Phys. Rev. B ${\bf 88}$, 155405 (2013).
%
\bibitem{araujo14}   M. A. N. Araújo, E. V. Castro, J. Phys.: Condens. Matter ${\bf 26}$, 075501 (2014).
%


\end{thebibliography}
\end{document}